\documentstyle[psfig]{mn}

\title{Spiral structure in the accretion disc of the binary IP~Pegasi}

\author[D. Steeghs et al.]
       {D. Steeghs, E.T. Harlaftis\thanks{Previous Address: Royal Greenwich Observatory, Apartado de Correos 321, E-38780 Santa Cruz de La Palma, Spain.} and Keith Horne\\
        Physics \& Astronomy, University of St Andrews, North Haugh, 
St Andrews, Fife KY16 9SS ({\tt ds10,ehh,kdh1@st-and.ac.uk})\\}

\date{Accepted 1997 July 22,Received ;in original form}
\pagerange{\pageref{firstpage}--\pageref{lastpage}}
\pubyear{1997}

\begin{document}


\maketitle
\label{firstpage}

\begin{abstract}

We have  found the first convincing evidence   for spiral structure in
the accretion disc of a close binary.  The eclipsing dwarf nova binary
IP Peg,  observed during the  end phase of  a  rise to outburst, shows
strong    Balmer and  Helium   emission  lines  in  its  spectra, with
asymmetric double  peaked velocity profiles  produced in the accretion
disc around the  white dwarf. To reveal  the  two armed spiral  on the
accretion disc, we de-project the observed emission line profiles onto
a  Doppler coordinate frame, a technique  known as Doppler tomography.
The two armed   spiral structure we  see in  the Doppler tomograms  is
expected to form when the disc becomes  sufficiently large in outburst
so that the tides  induced by the secondary star  can excite waves  in
the  outer disc. Such spiral waves  have  been predicted in studies of
tidal effects  in  discs and are   fundamental in  understanding  the
angular momentum budget of accretion discs.

\end{abstract}


\section{Introduction}

IP Pegasi  is  an interacting binary system   containing a white dwarf
receiving mass through  an accretion  disc  from a Roche lobe  filling
late type star.  These accretion disc  fed systems  called cataclysmic
variables (see Warner (1995) for an excellent overview) provide one of
the best laboratories for accretion physics due to their proximity and
convenient time  scales.  The strong  emission lines in  their spectra
originate in the accretion flow and  are powerful observational probes
of the    local  gas conditions.    The picture  of   a  viscous disc,
transporting  angular   momentum outwards  as  material slowly spirals
inwards, forms  the basis of  our understanding of accretion  flows in
X-ray binaries and AGNs as well.

One of  the  main longstanding problems of  accretion   discs is their
angular   momentum transport  mechanisms.    In order  to sustain  the
observed mass transfer rates highly  efficient viscous processes  must
be available to transport the angular momentum outwards.  Although the
famous $\alpha$ prescription (Shakura \&  Sunyaev 1973), which  scales
the  effective viscosity by a   dimensionless parameter $\alpha$,  has
been very succesfull  it  also shows  how  poorly these processes  are
understood.  Turbulent     magnetic fields  (Tout    \&  Pringle 1992,
Schramkowski   \& Torkelsson 1996) and  spiral   shocks (Spruit et al.
1987) are  two  promising mechanisms even  though  the effective $\alpha$ expected
from such models is still low. A second  issue that has received less
attention is the removal of the angular momentum at the outer disc via
a  tidal torque  between disc and  companion  star (e.g. Papaloizou \&
Pringle 1977).

IP  Peg is a member of   the subclass of CVs   called dwarf novae that
display  semi-periodic outbursts during  which the system brightens by
several magnitudes  as more mass   is suddenly transfered  through the
disc. These systems  provide  a great test   case for  accretion  disc
models.  IP  Peg is one of the  few  eclipsing dwarf novae,  where the
inclination of the  orbital plane ($\sim$80$^{\circ}$) is large enough
for the 0.5  M$_{\odot}$ companion star to  cover the 1.02 M$_{\odot}$
white dwarf and most of the accretion disc as it passes in front every
3.8 hours. IP Peg's outbursts have  an amplitude of about 2 magnitudes
and recur  roughly every 3 months during  which the  accretion disc is
the dominant light source.

We  present spectrophotometric observations of  the dwarf nova IP Peg
at the late stages  of a rise to outburst  and use Doppler  imaging to
map  the accretion disc.   Observations   are presented  in section  2
followed  by  the analysis of  the tomograms  in  section 3. The tidal
origin of the spirals is discussed in section 4.

\section{Observations}

The data we present here  are part of a long  term service program  to
study IP  Peg    throughout its outburst cycle.    Time-resolved   CCD
spectrophotometry with the 2.5m Isaac Newton Telescope on La Palma was
used to study  the strong emission lines  originating in the accretion
disc both  during quiescence  and  outburst.  Here we will   focus our
attention on the  data obtained during the  night of 19 August,  1993.
IP Peg had just gone into outburst  a day before and  was close to its
maximum brightness level. The Intermediate Dispersion Spectrograph was
used to obtain spectra between  6300 and 6800 \AA, covering  H$\alpha$
and HeI($\lambda 6678$) at a mean dispersion of 0.56 \AA\ pixel$^{-1}$
or 38 km s$^{-1}$pixel$^{-1}$.  A  1024$\times$1024-pixel TEK CCD chip
recorded long slit spectra of IP Peg and  a comparison star to account
for    slit-losses.   Neon arc   spectra  were  regularly recorded for
wavelength  calibration and the  flux standard BD+28$^{\circ}$4211 was
used for flux calibration. This setup allowed  us to optimally extract
spectra with an  absolute flux scale.  A  total of 15  spectra with an
exposure time of  360 s where  obtained sampling 60\%  of the 3.8 hour
binary orbit.  

The top  panels of  Figure 1   show the H$\alpha$  and  HeI(6678) line
profiles as a  function of binary phase  after subtracting a low order
spline fit to the continuum of  the individual spectra. Orbital phases
were calculated using the Wolf et  al.  (1993) ephemeris without their
quadratic term; 
\[ T_0(HJD)=2445615.4156  +  0.15820616 E \]

\noindent  with   $T_0$
corresponding   to   mid-eclipse.    The  AB$\sim$12.6  mag  continuum
increasing by $\sim 7\%$ during the 2 hour observing window, shows that
IP Peg was near the top of its rise to outburst, which typically lasts
1--1.5 d.

\section{Doppler Maps}

To  interpret the phase  dependent line profiles $f(v,\phi)$ (Fig. 1),
we     use Doppler tomography   (Marsh \&   Horne   1988), an indirect
de-projection technique very similar  to CAT scanning used in  medical
imaging. The Doppler map $I$(V$_x$,V$_y)$  gives the emission line flux
of gas moving with velocity vector $V=(V_x,V_y)$ in the rotating frame
of  the binary. As  the binary  rotates,  projections of  the rotating
velocity  vector onto the line  of sight traces  the sinusoidal radial
velocity curve;

\[ V(\phi)=-V_x \cos\phi + V_y \sin\phi   \]

\noindent 
The observed  line profiles $f(v,\phi)$  can  therefore be modelled as
projections  of   the map   $I$(V$_x$,V$_y)$  without  making  specific
assumptions about the form of the velocity field of the accretion flow
(see also Robinson, Marsh \&  Smak 1993 and  Horne 1991). 
A maximum    entropy
implementation  was  used where   the   Doppler  image  is  built   up
iteratively. Any given map is  projected to produce the predicted line
profiles  for  the  particular  map.  $\chi^2$  statistic  is  used to
determine goodness of fit while the entropy is maximised to select the
simplest image that can fit the data to the required $\chi^2$ value.

This technique  assumes that the disc  pattern is  constant throughout
the data set (in the co-rotating frame of the binary) so that the line
variations  can be modeled   by projection effects. Transient features
will therefore be  averaged   out over the   map so  that  the average
co-rotating  pattern is recovered.  Tidal  distortions co-rotate in the
binary frame and therefore do not suffer  from this restriction and are
ideally  recovered by   Doppler tomography. A  second  problem  can be
secular variability  of  the  system  within the    data set  used  for
tomography.  In  our case the continuum  showed little increase during
the course of our observations  (i.e.  outburst was developed) and  as
our  observations  cover   only  $\sim$ 2h,  which   is  sufficient to
calculate a Doppler image as more  than half of  the orbital period is
covered,   secular changes    were negligible.  Furthermore, line flux
variations  were compatible   with  the changing  contribution  of the
companion star as  the illuminated inner face  comes  into view, while
the disc contribution was stable.

Middle panels  of Figure 1  show the  two  maps constructed  from  the
observed H$\alpha$  and HeI(6678) line flux.   As a comparison, bottom
panels show predicted  data and  can be used  to  check  how well  the
Doppler  image reproduces  the  observed  line  emission. The gas   stream
trajectory and position of the companion  star's Roche lobe is plotted
based  on the known  system parameters (Marsh  \& Horne 1990).  Strong
secondary  star  emission (K$_2$=300 km s$^{-1}$)  is  visible in both
lines, a common feature of  dwarf novae in  outburst and is thought to
be due to irradiation of the inner face of the star. However, emission
from the companion has also been observed during quiescence (Harlaftis
et al. 1994) and can be related to intrinsic activity of the late type
star as the secondary star is co-rotating in a binary with a period of
only several hours. There is also a  weak low velocity component
in the H$\alpha$  image,  which was  observed a week  later by Steeghs
et. al (1996) who propose prominence like structures to be responsible
for this feature.  This emission  is   thus already present early   in
outburst, even though it is more pronounced a week later.

Disc  emission is centered on  the white dwarf (K$_1$=147 km s$^{-1}$)
and has a strong azimuthal asymmetry in the form of a two armed spiral
pattern.   Both lines show  similar structure  but  the arms are  more
sharply defined in the HeI map. The line flux in these spirals
is about a factor of  $\sim$4 stronger than  that of the disc emission
outside   the spirals pointing   to  considerable heating and  density
enhancement. The velocities of the disc material in
the two arms decrease from $\sim$700 km s$^{-1}$  down to $\sim$500 km
s$^{-1}$ with  increasing  azimuth, suggesting a highly  non Keplerian
flow.  A  Keplerian accretion  disc on the   other  hand would produce
circular rings   of   emission, each   velocity corresponding   to   a
particular Kepler radius ($V(r)=\sqrt{GM/r}$)  as has been observed in
tomographic studies of other binaries.  Note  that the two arms are  not perfectly symmetric, the
arm in the upper right of the tomogram is slightly stronger.

\begin{figure*}
\centerline{\psfig{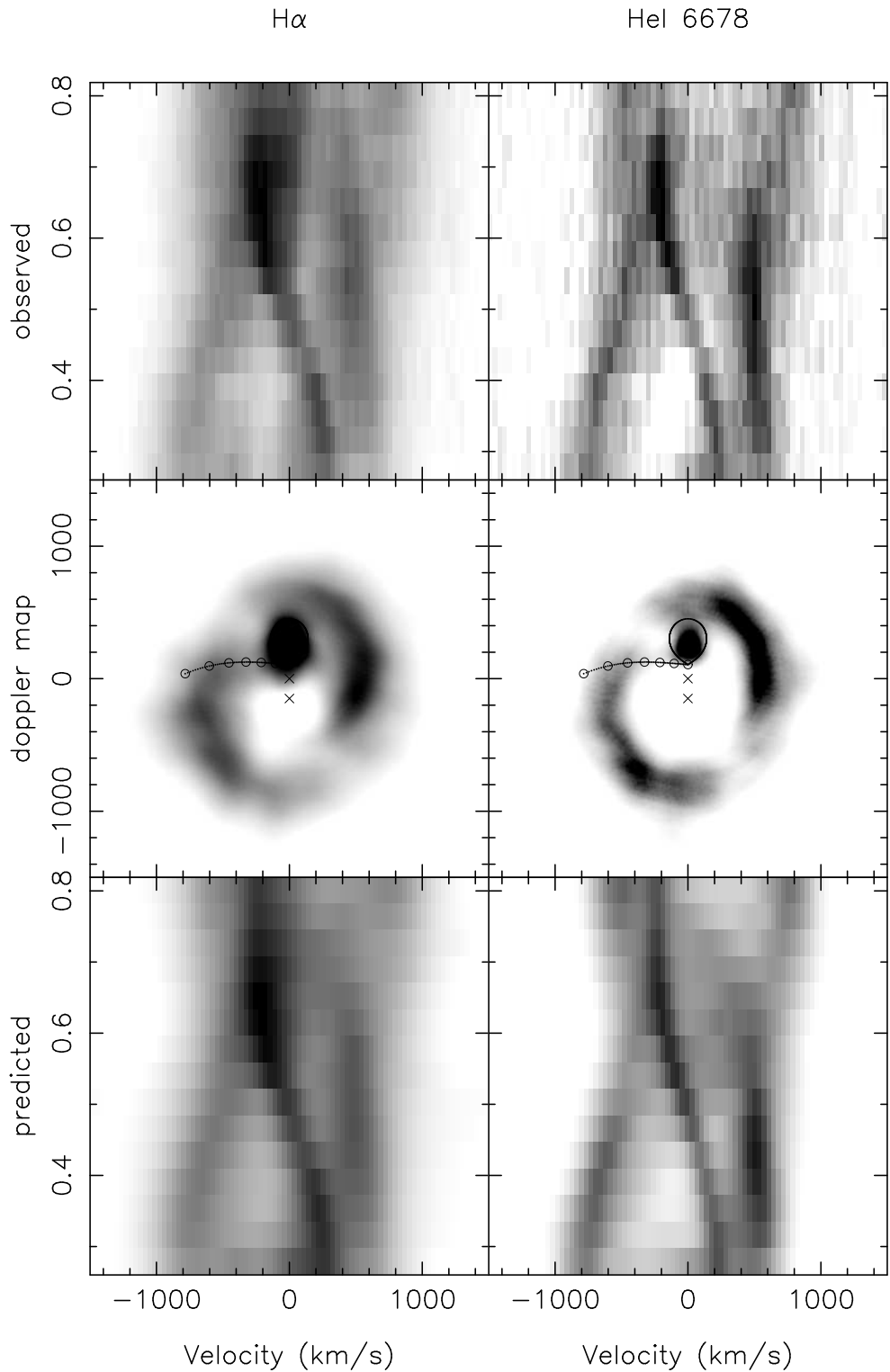}}
\caption{
Top  panels show the observed line  flux from IP Peg  as a function of
binary  phase  with    H$\alpha$   on the  left,    HeI(6678)  on  the
right.   Middle   panels  are    constructed Doppler   tomograms  with
theoretical  gas stream and  Roche  lobe plotted for comparison. Bottom
cross denotes  white dwarf, middle cross  the system center of mass at
V=(0,0). Bottom panels show  predicted data constructed by  projecting
the Doppler image  at the observed phases used  to determine  how well
the image fits our data.}
\end{figure*}

\section{Tides in the outer disc}

The presence of the companion star will perturb the disc material from
their   circular  Keplerian orbits  in    the  outer disc,  ultimately
resulting in intersecting orbits outside the radius referred to as the
tidal radius  (Paczynski 1977). For IP  Peg  this occurs  at $\sim$0.7
$R_{L_1}$  and  is  thought  to represent  the    maximum radius  of a
quiescent disc.  This tidal   interaction is essential in extracting  the
angular momentum,  transported outwards   through the disc  by  viscous
processes, from the disc via  a tidal torque. Hydrodynamic  simulations
(Sawada et al.   1986,  Savonije et  al.   1994, Heemskerk  1994)  and
analytical work (Spruit et al. 1987, Dgani et al.  1992) on this tidal
interaction show that  spiral waves, and  even shocks, are expected to
be generated  in   the accretion  disc   down  to quite   small radii,
depending on the Mach number of the disc flow. For hot accretion discs
(low  Mach numbers), these  trailing  waves can provide  a steady mass
transfer rate by   transporting angular momentum  outwards without the
need of intrinsic disc viscosity.  For the  high Mach numbers expected
in CV discs, the effective $\alpha$ is low, however ($\leq$ 0.01), and
is therefore likely not the dominant  transport mechanism in the inner
disc, but will still dominate the dynamics of the outer disc.

Many Doppler maps  have previously been constructed from  observations
of discs,  but those have never shown  obvious evidence for the spiral
waves predicted  by theory.  Our observations  now  for the first time
provide observational evidence  for a two  armed trailing spiral in  a
dwarf novae disc.   To confirm whether a two  armed  spiral can indeed
produce the observed line profiles, we constructed  a Doppler map of a
model disc containing two symmetric trailing  spiral arms, as shown in
Figure 2.  This  model assumes a two-armed  trailing spiral pattern in
the  spatial line emissivity of  the disc, covering  the outer part of
the disc between   0.4 and  0.9  $R_{L_1}$ (Figure  2,  bottom).   The
velocity coordinates conserve   its  azimuthal shape, resulting in   a
model Doppler image with two spirals as well (Figure 2, middle panel).
Note that the model  was optimised to  reproduce the velocities of the
observed spirals. The arms span $\sim 110^{\circ}$ in azimuth, and appear to be very open.  The quoted radii corresponding  to  this  are the
Kepler orbits that limit the spirals.
The predicted line profiles of this  model are shown  in the top panel
and demonstrates a  close resemblance to the  observed  data of Figure
1. The  key   signature   is  the   modulation of  the    double  peak
separation. The two peaks  measure the radial  velocity of material on
either side  of the disc moving  almost directly towards and away from
the  observer.  Their separation would be  constant  as a  function of
binary  phase for an axisymmetric  (Keplerian)  disc for example. Note
also the jump in velocity around phase  0.7 where one crosses from one
arm to the  other. While general  asymmetries in  the local emissivity
can be produced by non circular orbits, the fact that it has the shape
of a  spiral  strongly favors  the  interpretation that we  are indeed
seeing a spiral density  wave in the outer disc.   As the orbits start
to intersect pressure and viscous forces  will setup density waves and
possibly even shocks.

\begin{figure}
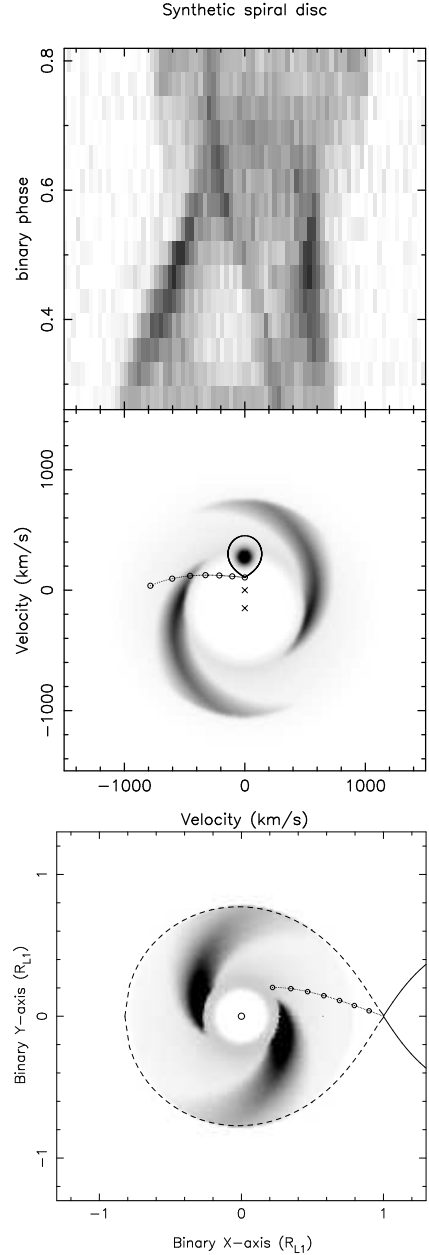

\centerline{\psfig{figure=fig2a.ps,width=5.5cm}}
\centerline{\psfig{figure=fig2b.ps,angle=-90,width=5.5cm}}
\caption{
A  model Doppler  tomogram   containing a  two armed  trailing  spiral
superposed  on  symmetric  disc   emission.  A Gaussian  spot  at  the
secondary is added to simulate  its contribution to the data.Top panel
shows the  predicted data from such a  system with the same  signal to
noise   as  our  observations   (compare  with top  panels   of Figure
1).   Middle panel is the  model  tomogram  and  bottom  panel shows a
spatial image of the disc emissivity pattern.}
\end{figure}

Tomography of the final stages of the outburst, about a week after our
data (Steeghs et al. 1996), reveals a similar asymmetry pattern in the
disc, most obviously in HeI. The much stronger companion star emission
dominates  over   the disc emission and   the  fainter  disc structure
suggests  the disc is shrinking and  the tidal distortions are damping
out.

Simulations suggest  large, hot discs  are  needed to  generate strong
waves  (Savonije et al. 1994).   Dwarf  novae  discs are  considerably
larger and hotter during outburst than in quiescence (e.g. Ichikawa \&
Osaki  1992, Wood et  al. 1989) due  to their high mass accretion rate
state.   Tidal forces  will   therefore  be  similarly enhanced.     A
combination of those two factors  (temperature and size) would explain
why quiescent discs do not  seem to show  such structure while (early)
outburst discs do.  Doppler   mapping studies of   dwarf novae in  the
early phase  of outburst on several  consecutive days, may be  able to
record the dynamical behaviour of the spiral waves.

The very  start of the outburst is  were the  two competing models for
the outburst, a  disc instability (DI) on one  hand (Osaki 1974) or  a
mass transfer burst (MTI) (Bath 1985)  on the other, predict different
disc behaviour. In the  MTI model, the  sudden addition of low angular
momentum gas causes the disc to shrink initially before it grows again
through viscous forces.  In the DI model, the disc  expands as soon as
it switches to the high viscosity  state at the  onset of the outburst
(e.g. Ichikawa   \& Osaki 1992).   Our data   suggests a large (almost
filling the full  Roche lobe), non  Keplerian accretion disc, possibly
exceeding its tidal radius, is present very  early on in the outburst,
and therefore favors a DI as the trigger of the outburst.

\section{summary}

 The tidal interaction manifestated in the spiral pattern turns out to
 be an important factor  for outburst discs.   Work is now in progress
 to   use different  observations of   this   phenomenon in  different
 emission lines    and at  different  epochs   to  sample the physical
 conditions of the disc material. Observing high ionization lines like
 HeII can  show   the presence  of   shocks   and  will indicate   the
 implication for   the  angular momentum budget.    Furthermore future
 observations of disc  structure in different  objects (with different
 mass ratios and disc sizes) will provide us with a new insight in
 tidal theory  and perhaps the  outburst mechanism.  In this way dwarf
 novae disc provide an excellent laboratory for tides in astrophysical
 discs, since the   time scales  of the outbursts  lasting   a week and
 recurring every couple of  months, allows one  to study the dynamical
 behaviour of  the  disc  and its tidal    response. Tidal spirals  in
 galaxies for example, thought to be generated in the same manner by a
 companion galaxy, have  very  long  dynamical  time scales making   it
 impossible to study their evolution.

\section*{Acknowledgments}

We thank Tom Marsh for his  valuable support in Doppler tomography and
Henk Spruit for  fruitful discussion.  The  Isaac Newton  Telescope is
operated on La  Palma by the Isaac  Newton Group of telescopes,  Royal
Observatories  in the Spanish  Observatorio del Roque de los Muchachos
of the Instituto de Astrofisico de Canarias.

\appendix

\bsp

\label{lastpage}

\end{document}